\documentclass[prl,amsmath,superscriptaddress,twocolumn,showpacs]{revtex4}
\usepackage{epsfig}

\usepackage{amssymb}
\usepackage{amsmath}
\usepackage{amsfonts}
\usepackage{bbold}
\usepackage{bm}
\usepackage{graphicx}
\usepackage{xcolor}

\newcommand{\nn}{\nonumber}
\newcommand{\ud}{{\textrm{d}}}

\begin{document}
\title{Time Evolution of Many-Body Localized Systems with the Flow Equation Approach }

\author{S. J. Thomson}
\author{M. Schir\'o}
\affiliation{Institut de Physique Th\'{e}orique, Universit\'{e} Paris Saclay, CNRS, CEA, F-91191 Gif-sur-Yvette, France}
\date{\today}

\begin{abstract}
The interplay between interactions and quenched disorder can result in rich dynamical quantum phenomena far from equilibrium, particularly when many-body localization prevents the system from full thermalization. With the aim of tackling this interesting regime, here we develop a semi-analytical flow equation approach to study time evolution of strongly disordered interacting quantum systems.  We apply this technique to a prototype model of interacting spinless fermions in a random on-site potential in both one and two dimensions. Key results include (i) an explicit construction of the local integrals of motion that characterize the many-body localized phase in one dimension, ultimately connecting the microscopic model to phenomenological descriptions, (ii) calculation of these quantities for the first time in two dimensions, and (iii) an investigation of the real-time dynamics in the localized phase which reveals the crucial role of $l$-bit interactions for enhancing dephasing and relaxation. 
\end{abstract}

\maketitle

\textit{Introduction - } Quenched random disorder can have dramatic effects on transport and dynamical properties of quantum many-body systems, leading to a complete breakdown of diffusion and to the localized insulating behavior of a non-interacting quantum particle in a random external potential~\cite{AndersonPR58}. Theoretical investigations in the past decade have shown that localization effects can survive in the presence of many-body interactions and finite temperature (or finite energy density) in closed isolated quantum systems~\cite{FleishmanAndersonPRB80,GornyiEtAlPRL05,BaskoAleinerAltshulerAnnPhys06,OganesyanHusePRB07,ZnidaricEtAlPRB08,
MonthusGarelPRB10,Pal_Huse_PRB2010,ImbrieJStPh2016,HuseReviewMBL}, a surprising result with profound consequences for our basic understanding of quantum statistical mechanics~\cite{BaskoAleinerAltshulerPRB07,AleinerAltshulerShlyapnikovNatPhys10,HuseEtAlPRB13,KhemaniEtAlPRL16}. While the interplay of interaction and disorder on the equilibrium low temperature physics of quantum systems has been thoroughly studied~\cite{LeeRamakrishnanRMP85}, these recent developments have triggered a new wave of interest into the problem and cast it into a completely new light, one where dynamics and out of equilibrium phenomena play a key role.

Very recently, the first signatures of many-body localization (MBL) have been observed with a number of experimental platforms, including ultra-cold atoms in quasirandom optical lattices~\cite{BlochMBL2015}, ion traps with programmable random disorder~\cite{SmithEtAlNatPhys16} and dipolar systems made by nuclear spins~\cite{KaiserMBL2015,KucskoEtAl_arxiv16}. 
The theoretical properties of MBL have mostly been discussed in the context of random quantum spin models~\cite{FisherPRB95} (or equivalently, interacting random spinless fermions), and mostly in one dimension. In particular, the dynamics of entanglement and its structure in highly excited MBL eigenstates has been largely unveiled~\cite{Bardarson_Pollmann_PRL12,SerbynPapicAbaninPRL13,VoskAltmanPRL13} and a phenomenological description of fully MBL phases has been proposed~\cite{SerbynPapicAbaninPRL13_2,HuseNandkishoreOganesyanPRB14} in terms of an extensive set of \emph{emergent} local integrals of motion which are conserved by the unitary dynamics, thus preventing complete thermalization. Such local degrees of freedom (also called localized bits or $l$-bits) are smoothly connected to those of the non-interacting Anderson insulator through a quasi-local unitary transformation~\cite{KimEtAl_arxiv14,Ros+15,Rademaker+16,MonthusJStatMech16,ImbrieRosScardicchioAnnPhy17,Monthus_Arxiv17,KulshreshthaEtAl_arxiv17,GoihlEtAl_arxiv17},  thus suggesting an underlying concept of adiabaticity similar, to a certain extent, to the Landau Fermi Liquid construction~\cite{BeraEtAlPRL15}. Further consequences of this emergent integrability have been discussed, specifically concerning broken symmetry phases~\cite{HuseEtAlPRB13}, memory of initial conditions and quantum coherence~\cite{VasseurEtAlPRB14,BahriEtAlNatComm15}.
More recently the focus has moved toward understanding the properties of the transition~\cite{KhemaniEtAlPRX17} from MBL into the fully ergodic regime, the anomalies appearing on both side of the transition~\cite{BarLevEtAlPRL15,AgarwalEtAlPRL15,VoskHuseAltmanPRX15,PotterVasseurParameswaranPRX15,GopalakrishnanPRB15,GopalarishnanEtAlPRB16,
LuitzBarLevPRL16,ZnidaricEtAlPRL16} or the effect of periodic driving~\cite{PonteEtAlPRL15,LazaridesEtAlPRL15}  (see Ref.~[\onlinecite{BardarsonEtAlAnnderPhysik2017}] for a recent topical review).

From the theoretical point of view, the majority of results in the literature have been obtained with numerical approaches, in most cases exact diagonalization or matrix product state simulations. It is therefore quite urgent to develop alternative methods which can provide analytical insights on the MBL phenomenon and its properties, particularly those related to dynamics out of equilibrium.

With this aim, in this work we present a novel approach for studying time evolution of interacting, strongly disordered quantum many-body system.  An extension of the established Flow Equation (FE) method traditionally used for translationally invariant (“clean”) systems both in and out of equilibrium~\cite{Kehrein07,Kehrein_PRB_08,Hackl_floweq09} this approach builds a series of continuous unitary transformations (CUTs) to iteratively diagonalise the Hamiltonian of the system in real-space, for a given realization of disorder. First attempts in this direction have appeared recently both in the context of non-interacting disordered quantum systems~\cite{Quito+16,RoyDasPRB15} and for genuine MBL problems, where the FE approach has been formulated as an exact scheme~\cite{Monthus16} and implemented on random quantum spin chains as numerical algorithm to diagonalize the Hamiltonian matrix in the full Hilbert space~\cite{Pekker+16,SavitzRefael_arxiv17}. 
These \emph{exact} implementations, although very powerful in principle, remain limited to rather small system sizes. Here, we instead proceed differently and introduce a semi-analytical version of the FE approach inspired by the large literature on CUTs for interacting quantum many-body systems~\cite{HeidbrinkUhrigPRL01,Kehrein07,Kehrein_prl08,HamerlaEtAlPRB10,KronesUhrigPRB15}. 

As we will show, this method allows us to address both static and - more importantly - dynamical properties of large disordered quantum many-body systems in the MBL phase, a problem so far studied mainly with numerical methods~\cite{KhatamiEtAl_PRE12,SerbynPapicAbaninPRB14,BarLevReichmanPRB14,MondainiRigolPRA15}, and even to discuss MBL physics in two dimensions, something which is currently out of reach for most state of the art approaches. A related discrete displacement transform has also been used to study MBL, but limited to static applications~\cite{Rademaker+16,Rademaker+17}. 

\textit{Disordered Lattice Fermions and the Flow Equation Approach - }  In order to describe the method and discuss our results, we will focus on a lattice model of disordered, interacting, spinless fermions related via the Jordan-Wigner transformation to the XXZ quantum spin chain model in random field which has been extensively studied in the context of MBL. Its Hamiltonian reads:
\begin{align}\label{eqn:Hxxz}
\mathcal{H} = \sum_i h_i c^{\dagger}_i c_i + \Delta\sum_{i}  n_i\,n_{i+1}+
J\sum_{i}  (c^{\dagger}_i c_{i+1} +hc)
\end{align}
with $n_i=c^{\dagger}_ic_i$, where the on-site random field is drawn from a box distribution $h_i \in [-W,W]$ and we set $J=1/2$ as our unit of energy, to map exactly on the XXZ spin chain after Jordan-Wigner. Notice that FEs are a rather general and flexible approach which can be applied to other quantum disordered problems (see discussion for future applications). The basic idea of the FE approach is to iteratively diagonalize the Hamiltonian of the system by a CUT $U(l)$ parametrized by a scale $l$ and generated by an anti-Hermitian operator $\eta(l)$, such that $U(l)=T_l \exp \left( \int \eta(l) \ud l \right)$. The flow of any operator $O(l)$ under this transform is given by $dO/dl = [\eta(l),O(l)]$. When applied to the Hamiltonian, this is spiritually similar to a standard renormalization group treatment, where the `fixed point' in the $l \to \infty$ limit is a diagonal Hamiltonian with renormalized couplings. The generator $\eta(l)$ is itself scale-dependent and changes throughout the flow. Here we use Wegner's choice for the generator and choose it to be the commutator of the (scale-dependent) diagonal and off-diagonal parts of the Hamiltonian, $\eta(l)=[\mathcal{H}_0(l),V(l)]$. This choice guarantees that the off-diagonal terms vanish in the $l \to \infty$ limit; other choices of generator are also possible~\cite{Monthus16}. An exact parametrization of this flow requires, for a generic interacting quantum many-body problem, either a large (exponential in size) number of matrix elements~\cite{Pekker+16} or running couplings~\cite{Monthus16} and is therefore limited to rather small systems . The key idea of our approach is to take advantage of the insights on the MBL phase to parameterise this diagonalisation flow in terms of a few “relevant” operators that most closely describe the fixed point Hamiltonian. This amounts to making an ansatz for the \emph{running} Hamiltonian, which we choose to be:
\begin{align}\label{eqn:Hansatz}
\mathcal{H}(l) &= \sum_i h_i(l) :c^{\dagger}_i c_i: + \sum_{ij} \Delta_{ij}(l) :c^{\dagger}_{i} c_{i} c^{\dagger}_{j} c_{j}: \\
&\quad \quad + \frac{1}{2}\sum_{ij} J_{ij}(l) (:c^{\dagger}_i c_{j}: + :c^{\dagger}_j c_i:)\equiv \mathcal{H}_0(l)+V(l), \nn
\end{align}
and disregard all newly generated terms outside this \emph{variational} manifold. While this approximation makes the decay of the off diagonal terms no longer guaranteed \emph{a priori}, we argue that the resulting error can be kept under control, particularly in the localised phase~\cite{SM_FlowEq}. A few comments are in order, concerning the above ansatz. Firstly, with the aim of targeting the MBL phase, we have chosen the first non-trivial terms responsible for pairwise interactions among $l$-bits, while higher order (diagonal or off-diagonal) terms have been discarded. These can in principle be accounted for at any order of the ansatz, at the cost of increasing the number of running couplings and the complexity of evaluating the flow equations. We expect this choice to be valid deep with in the MBL phase, but to break down in the delocalised phase (see below).
Secondly, we have adopted the normal-ordering notation $:\hat{O}:$ with respect to an initial product state  (see Refs.~ [\onlinecite{SM_FlowEq,Kehrein07}]) in such a way to (i) fix unambigously the precise form of the flow equations and (ii) enhance the convergence properties of the truncation scheme~\cite{Kehrein07} . The flow equations for the running couplings can be obtained from evaluating $\ud \mathcal{H} / \ud l = [\eta(l), \mathcal{H}(l)]$, with $\eta(l)=[\mathcal{H}_0(l),V(l)]$, and after a lengthy but otherwise straightforward calculation are given by~\cite{SM_FlowEq}:
\begin{align}
\frac{\ud h_i}{\ud l} &= 2 \sum_{j} J_{ij}^{2} (h_i - h_j) \label{eq.hflow} \\
\frac{\ud J_{ij}}{\ud l} &= -J_{ij} (h_i - h_j)^{2} - \sum_{k} J_{ik} J_{kj} (2h_k - h_i - h_j)+ \nn\\
& -8 \sum_k J_{ij} (\Delta_{ik} - \Delta_{jk})  (C_{ki}\Delta_{ik} - C_{kj} \Delta_{jk}) +\nn\\
& -8 J_{ij} \Delta_{ij}^2 (C_{ij}+C_{ji}) \label{eq.jflow} \\
\frac{\ud \Delta_{ij}}{\ud l} &= 4 \sum_{k} \left[ J_{ik}^{2} (\Delta_{ij} - \Delta_{kj}) + J_{jk}^{2} (\Delta_{ij} - \Delta_{ki}) \right]- 8 J_{ij}^{2} \Delta_{ij} \label{eq.deltaflow}
\end{align}
where we have defined $C_{ij}=\langle n_i\rangle^2 \langle n_j\rangle$. In practice, we numerically solve this flow, starting from the microscopic initial conditions $h_i(0)=h_i,J_{ij}(0)=J\delta_{i,i+1},\Delta_{ij}(0)=\Delta\delta_{i,i+1}$  up to some large but finite value of $l$ where the off-diagonal elements have decayed to required accuracy and we are left with a diagonal Hamiltonian. 

\begin{figure}[t]
\begin{center}
\includegraphics[width= 1.\linewidth]{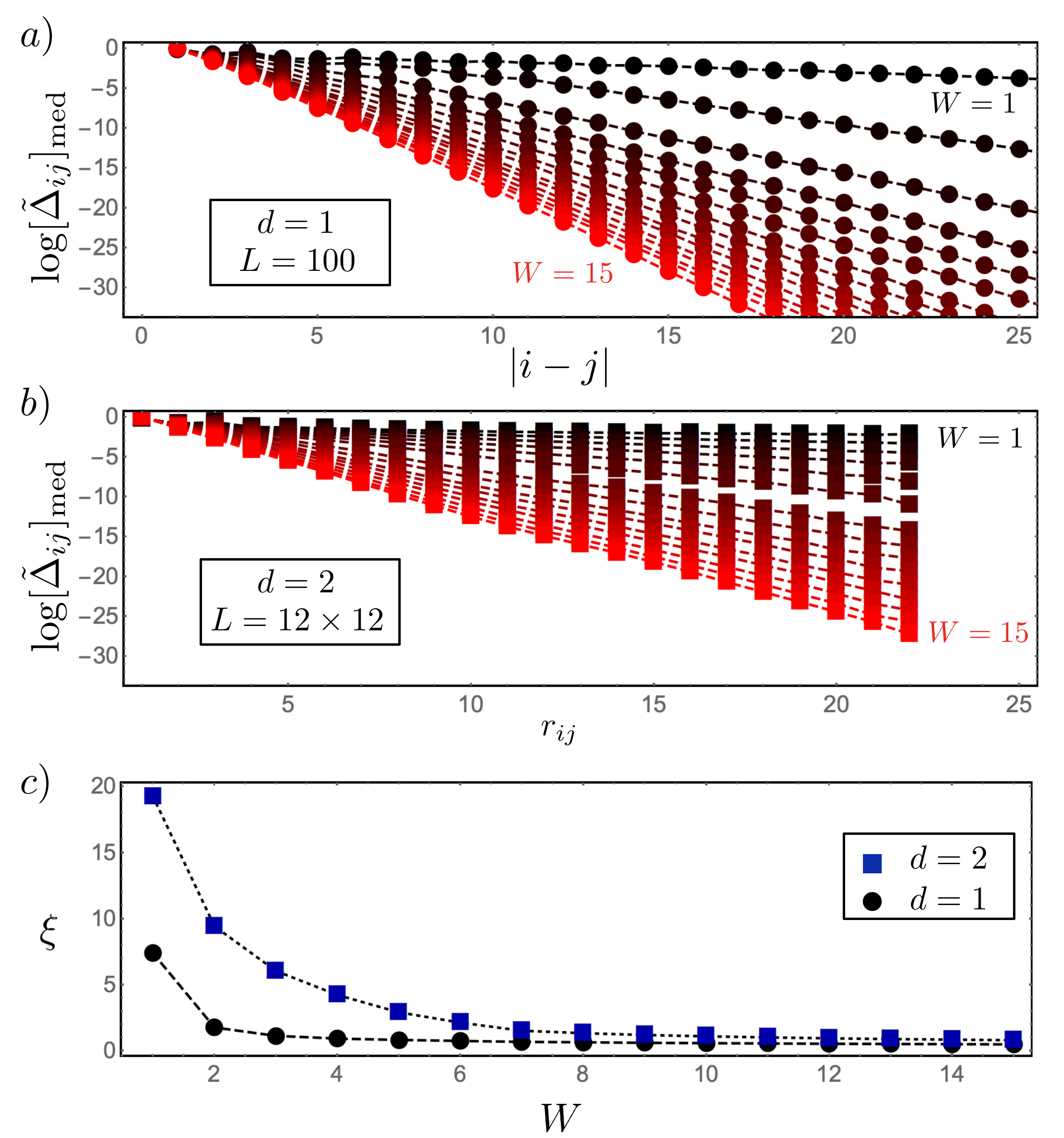}
\caption{a) Real-space exponential decay of fixed-point couplings $\tilde{\Delta}_{ij}$, describing mutual interactions among $l$-bits, for increasing values of disorder (top to bottom) in $d=1$. b) The same quantity for $d=2$, plotted to the same scale. c) \emph{Localization } length scale extracted from   $\tilde{\Delta}_{ij}$ (see main text) as a function of disorder from large to small. System size is $L=100$ for $d=1$ and $L=12\times12$ for $d=2$, and the number of disorder realizations for each is $N_{S}=100$.}\label{fig:fig1}
\end{center}
\end{figure}
\textit{Benchmark: Properties of the MBL Phase with FE - } To assess the validity of our ansatz~(\ref{eqn:Hansatz}) we compute the spectrum using the FE, which simply amounts to read off the eigenvalues of the diagonal fixed point Hamiltonian $\mathcal{H}(\infty)\equiv  \sum_i \tilde{h}_i n_i +\sum_{ij} \tilde{\Delta}_{ij} n_i n_j$, with 
$\tilde{h}_i\equiv h_i(\infty)$ and $\tilde{\Delta}_{ij}\equiv \Delta_{ij}(\infty)$ and compare them with exact diagonalisation results for a chain of length $L=12$. Such analysis~\cite{SM_FlowEq} confirms that the agreement is excellent in the strongly localized phase down to disorder $W\sim 4$ and deteriorates upon entering the delocalized phase.
As already mentioned, a natural outcome of our approach is the explicit construction of an effective Hamiltonian for the $l$-bits of the system. It is therefore interesting to discuss the structure of these degrees of freedom and their mutual interaction. In Figure~\ref{fig:fig1}a) we plot $\tilde{\Delta}_{ij} = \tilde{\Delta}(|i-j|)$ vs the length $|i-j|$ for different disorder strengths. In all cases we find that the couplings $\tilde{\Delta}_{ij}$ decay exponentially with distance, however the decay is much slower in the small-disorder regime. We can extract a localisation length $\xi$ by fitting these with a decay of the form $\tilde{\Delta}_{ij} \propto \exp{(-|i-j|/\xi)}$. Plotting this localization length against disorder strength, we see that it increases with decreasing disorder, however as in Ref.~\cite{Rademaker+17} there is no sign of a delocalisation transition (known to occur around  $W_c \approx 3.5$~\cite{LuitzEtAlPRB15}) in this averaged value of $\tilde{\Delta}_{ij}$. It is likely that the higher-order terms not included in our ansatz~(\ref{eqn:Hansatz}) become relevant near the transition and would need to be included to capture it. We have also computed~\cite{SM_FlowEq} the full probability distributions of the $\tilde{\Delta}_{ij}$ and find approximate power-law decay at all disorder strengths, consistent with both Refs. [\onlinecite{Pekker+16,Rademaker+17}] in the regime of disorder and lengthscale where our method is accurate.

\textit{Localization in Two Dimensions - } A major advantage of our truncated FE approach is that it can be easily extended to address the fate of MBL beyond one dimension, an issue which has so far remained largely unexplored~\cite{BarLevReichmanEPL16,ChandranEtAlPRB16} despite its experimental relevance~\cite{ChoiEtAlScience16}. Within our implementation, information about lattice geometry and dimensionality enters only in the initial condition, while the flow of the running couplings remains unchanged and is still given by Eqs. (\ref{eq.hflow}$-$\ref{eq.deltaflow}). Therefore the previous analysis can be straightforwardly extended to two dimensions by making the appropriate modifications to the initial Hamiltonian. We measure the distance on the $d=2$ lattice by the Manhattan distance, i.e. $r_{ij}=|x_i-x_j|+|y_i - y_j|$ where $(x_i,y_i)$ are lattice co-ordinates. The fixed-point couplings are shown in Fig. \ref{fig:fig1}b): the system is significantly less localized than in $d=1$. The localization length extracted from the lowest disorder strength is larger than our system size, possibly indicating that the system is delocalized at small disorder, with a clear shift towards fast exponential localization of the $d=2$ ``$l$-bits'' at strong disorder. Whether there is a true phase transition between these is left for future work, and the same caveats of our truncated Hamiltonian in $d=1$ apply to the $d=2$ system.

\textit{Time Evolution in the MBL Phase with Flow Equations -  }While the FE method provides a natural framework to understand the local integrals of motion picture and for computing static properties of the MBL phase, another significant advantage is its potential to investigate time evolution and dynamics of strongly disordered interacting quantum systems. 
The key observation is that time evolution becomes trivial in the basis where Hamiltonian is diagonal. The challenging part is to keep track of the change of basis, which is what the FE naturally does. More formally, any given time dependent average of the form $O(t)=\langle\Psi_0\vert e^{i\mathcal{H}t}Oe^{-i\mathcal{H}t}\vert\Psi_0\rangle$ can be also written as: 
\begin{align}\label{eqn:dynamics}
O(t)= \langle\Psi_0\vert U^{\dagger}(l) e^{i\mathcal{H}(l)t}O(l)e^{-i\mathcal{H}(l)t}U(l)\vert\Psi_0\rangle ,
\end{align}
an expression which is particularly useful for $l=\infty$, where it amounts to flowing the observable under $U(l)$, time evolving it with the diagonal $H(\infty)$, and flowing it back to the original basis before taking the expectation value~\cite{Kehrein_PRB_08,Hackl_floweq09}. Such a backward transformation needs to be done for each timestep $dt=T_{max}/N$ during the evolution up to time $T_{max}$, resulting in the solution of a large number of differential equations, $O(N\times L^2)$ for a system of size $L$, which represents the main computational challenge behind this approach.

As a concrete and non-trivial example, we study the dynamics of the model in Eq.~(\ref{eqn:Hxxz}) for a chain of length $L=64$ in $d=1$, starting from a product state with a charge density wave (CDW) pattern, i.e. $\vert\Psi_0\rangle=\vert .. 010101...\rangle$.  Following the experimental realization of MBL with cold-atoms in quasi-random disorder~\cite{BlochMBL2015} this kind of protocol has attracted considerable attention~\cite{LuitzEtALPRB16,BiroliTarzia_arxiv17}. We monitor the dynamics of the system by tracking the density in the middle of the chain. To do so, it is necessary to compute the flow of number operator $n_i$ under $U(l)$,  which is given as usual by $\frac{\ud n_i}{\ud l} = [\eta(l), n_i]$ with the same generator $\eta(l)=[\mathcal{H}_0(l),V(l)]$ used to diagonalize the Hamiltonian in Eq.~(\ref{eqn:Hxxz}). As with the Hamiltonian, the operator flow is not closed but rather generates successively higher order terms describing an effective operator dressing by the many-body processes. We parametrize this flow by means of an ansatz including a single-particle hole excitation, which would be exact in the case of  a non-interacting system~\cite{Quito+16} and which we expect to work well in the localised phase,
\begin{align}\label{eqn:density_ansatz}
n_i(l) &= \sum_j \alpha^{i}_j(l) :c^{\dagger}_j c_j: + \sum_{jk} \beta_{jk}(l) :c^{\dagger}_j c_k:,
\end{align}
A simple calculation for the flow of $\alpha^i_j,\beta_{jk}$ gives:
\begin{align}
\frac{\ud \alpha^i_j}{\ud l} &= -2 \sum_i J_{ij} (h_i - h_j) \beta_{ij}, \\
\frac{\ud \beta_{jk}}{\ud l} &= -J_{jk} (h_k-h_j)(\alpha^i_k-\alpha^i_j) \nn\\
&+\sum_{n} \left[ J_{nj} (h_j - h_n) \beta_{nk} + J_{nk} (h_n - h_k) \beta_{nj} \right], \label{eq.nflow}
\end{align}
which has to be solved with the initial conditions $\alpha^i_j(0)=\delta_{ij}$ and all $\beta_{jk}(0)=0$.  
\begin{figure}[t]
\begin{center}
\includegraphics[width= 1.\linewidth]{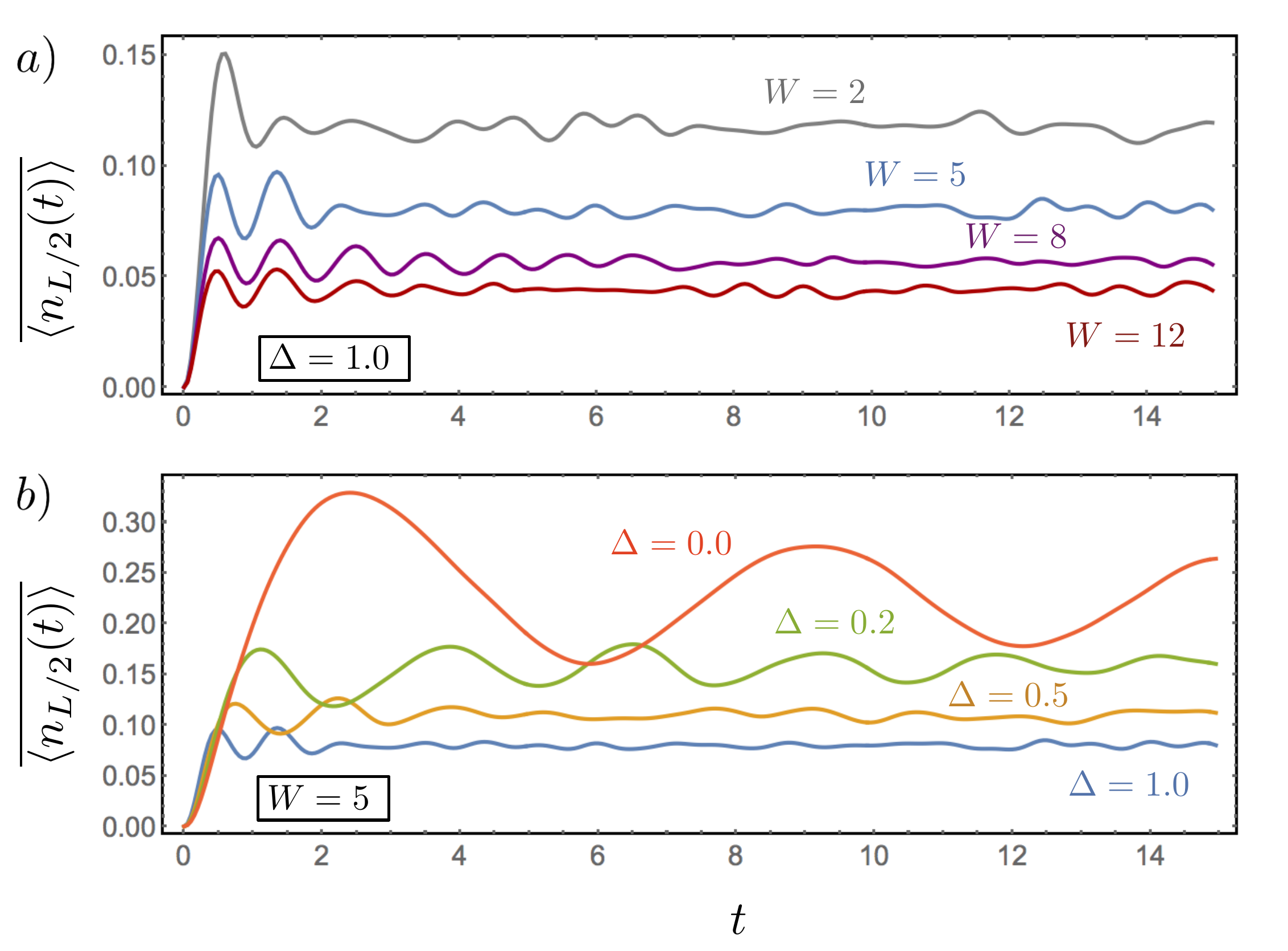}
\caption{Real-time evolution of the disorder-averaged density of fermions in the middle of the chain, starting from a CDW initial state. Top panel: Effect of disorder on the melting of CDW correlations, which suppresses the fluctuations but does not appear to change the relaxation timescale. The frequency of the oscillations at early times are set by the interaction strength. Bottom panel: Interaction effects between $l$-bits strongly quench their dynamics and induces decoherence.
Parameters: $L=64,N_S=500$.}\label{fig:fig2}
\end{center}
\end{figure}
We then have to time-evolve the density operator in Eq.~(\ref{eqn:density_ansatz}) with respect to the diagonal Hamiltonian $\mathcal{H}(\infty)=\sum_i \tilde{h}_i n_i +\sum_{ij} \tilde{\Delta}_{ij} n_i n_j$. 
Although diagonal, time evolution starting from a generic state (linear combination of many $l$-bit configurations) remains non-trivial to compute and essentially analagous to a classical statistical mechanical problem. We therefore resort to a decoupling of the equations of motion which is able to capture dephasing between different $l$-bits. A simple calculation gives the exact dynamics for the density $n_i$:
\begin{align}
i \frac{\ud n_i}{\ud t} &= \sum_{jk} \beta_{jk} (\tilde{h}_k - \tilde{h}_j) :c^{\dagger}_j c_k: \nn \\ 
& + 2 \sum_{kjm} \beta_{jk}(\tilde{\Delta}_{km} - \tilde{\Delta}_{jm}):c^{\dagger}_j c_k c^{\dagger}_m c_m:.
\end{align}
which can be decoupled using $:c^{\dagger}_j c_k c^{\dagger}_m c_m: \approx :c^{\dagger}_j c_k: \langle c^{\dagger}_m c_m \rangle$ to obtain a closed solution of the form:
\begin{align}
& n_i (l=\infty, t) = \sum_j \tilde{\alpha}^i_j n_j + \sum_{jk} \textrm{e} ^{i \phi_{jk} t } \tilde{\beta}_{jk} c^{\dagger}_j c_k,  \\
\phi_{jk} & = (\tilde{h}_k - \tilde{h}_j) + 2 \sum_{m} (\tilde{\Delta}_{km} - \tilde{\Delta}_{jm}) \langle n_m \rangle.
\end{align}
From this result we already see the crucial role played by interactions among $l$-bits in enhancing dephasing and relaxation. Finally, to obtain results for the density of physical fermions we need to transform back into the original basis, following Eq.~(\ref{eqn:dynamics}), which involves solving the backward flow using time-dependendent initial conditions  $\alpha^i_j(0,t)=\tilde{\alpha}^i_j $ and $\beta_{jk}(0,t)=\textrm{e} ^{i \phi_{jk} t } \tilde{\beta}_{jk}$.  In Fig.~\ref{fig:fig2} we plot the time evolution of the fermion density in the middle of a chain of length $L=64$, after averaging over $N_S=500$ disorder samples. We see that upon increasing disorder strength, the initial CDW pattern remains longer and longer lived, a signature of the enhanced memory of initial conditions typical of the MBL phase. Furthermore, the effect of interactions is also rather remarkable: it rapidly quenches the wide coherent oscillations of the Anderson insulator down to a stationary state with enhanced imbalance, a direct signature of how the $l$-bit interactions are able to induce dephasing.

\textit{Conclusions and Perspectives - } To conclude, in this work we have introduced an analytical flow equation approach to study disordered lattice fermion models and applied it to the MBL problem. The results we have presented show that the method is controlled in the interacting localized phase, while to capture the transition it is necessary to go beyond the ansatz Eq.~(\ref{eqn:Hansatz}). Nevertheless, the possibility to study the localized phase for large systems is rather appealing and immediately suggest a number of short and long term perspectives. One could use the FE approach to study quantum impurity problems in disordered environments, such as the central spin model~\cite{PonteEtAl_arxiv17} or other models of qubits  coupled to an MBL system~\cite{VasseurEtAlPRB14,vanNieuwenburgEtAlPRB16}, or to study non-linear transport in the MBL phase, along the lines of Ref.~\onlinecite{LangeEtAlNatComm17}.  Another intriguing perspective which seems worthy of pursuit is the application of the concept of CUTs and FEs to more general time-dependent problems, such adiabatic evolution or driven Floquet MBL problems~\cite{BordiaEtAlNatPhys17}, or to open and dissipative MBL problems described by a Lindblad master equation~\cite{MedvedyevaEtAlPRB15,LeviEtAlPRL16,LuschenEtAlPRX17} to study more exotic out-of-equilibrium effects in disordered quantum systems.

\emph{Acknownledgements - } We acknowledge discussions with  F. Alet, P. Crowley, S. Gopalakrishnan, C. Monthus, V. Oganesyan,  V. Ros. This work was supported  by a grant ``Investissements d'Avenir" from LabEx PALM (ANR-10-LABX-0039-PALM). 



\end{document}


\title{Supplementary Material to ``Time Evolution of Many Body Localized Systems with the Flow Equation Approach"}

\author{S. J. Thomson}
\author{M. Schir\'o}
\affiliation{Institut de Physique Th\'eorique, Universit\'e Paris-Saclay, CNRS, CEA, F-91191 Gif-sur-Yvette, France}
\date{\today}

\maketitle

This Supplementary Material is organized as follows: in section I we define normal ordering of operators, needed to derive Flow Equations for an interacting problem; in section II we give the explicit expression for the generator of the flow $\eta(l)$ starting from the ansatz given in the main text for the running Hamiltonian and discuss some qualitative features of the resulting flow. Section III is devoted to comparison with Exact Diagonalization results for the same model studied in the main text. Finally in section IV we present results for the full distribution of $l-$bit interactions.

\section{Normal-Ordering Convention}

Follow Ref. \onlinecite{Kehrein07}, we define the contractions of the operators using:
\begin{align}
\{ c^{\dagger}_{\alpha}, c_{\beta} \} &= G_{\alpha \beta} + \tilde{G}_{\beta \alpha}, \\
\textrm{with} \quad \quad G_{\alpha \beta} &= \langle c^{\dagger}_{\alpha} c_{\beta} \rangle, \\
\tilde{G}_{\alpha \beta} &= \langle c_{\alpha} c^{\dagger}_{\beta} \rangle,
\end{align}
Here, we shall make no assumptions as to the state with respect which we normal order and keep all normal-ordering corrections. Define the normal ordering procedure using the $:\hat{O}:$ notation as:
\begin{align}
: c^{\dagger}_{\alpha} c_{\beta} : &= c^{\dagger}_{\alpha} c_{\beta} - \langle c^{\dagger}_{\alpha} c_{\beta}\rangle \\
&= c^{\dagger}_{\alpha} c_{\beta} - G_{\alpha \beta} ,\\
: n_i n_j : &= :c^{\dagger}_i c_i c^{\dagger}_j c_j: = c^{\dagger}_i c_i c^{\dagger}_j c_j - G_{jj} n_i - G_{ii} n_j - \tilde{G}_{ij} c^{\dagger}_i c_j -G_{ij} c_i c^{\dagger}_j + G_{ii}G_{jj} + \tilde{G}_{ij}G_{ij},
\end{align}
where the first identity follows directly from Wick's theorem and the second is from Ref. \onlinecite{Kehrein07} but can be derived in the same way.

To calculate the commutation relations of normal-ordered strings of operators, we use the following theorem\cite{Kehrein07}:
\begin{align}
:O_1 (A): :O_2 (A'): = : \exp \left( \sum_{kl} G_{kl} \frac{\partial^2}{\partial A'_{l} \partial A_{k}} \right) O_{1} (A) O_{2}(A'):,
\end{align}
where $A$ and $A'$ are the set of labelled operators in the expression $O_1$ and $O_2$ which in our case are just strings of Fermi operators.

We can evaluate this by expanding the exponential - for simple expressions, the expansion can be exact, as all derivates are zero above some order greater than the number of operators in $O_1$ or $O_2$. For example, for the commutator of two quadratic terms:
\begin{align}
:c^{\dagger}_{\alpha} c_{\beta}: :c^{\dagger}_{\gamma} c_{\delta}: &=  : \left( 1 + G_{\alpha \delta} \frac{\partial^2}{\partial c^{\dagger}_{\alpha} \partial c_{\delta}} + \tilde{G}_{\beta \gamma} \frac{\partial^2}{\partial c_{\beta} \partial c^{\dagger}_{\gamma}} + G_{\alpha \delta} \tilde{G}_{\beta \gamma} \frac{\partial^4}{\partial c^{\dagger}_{\alpha} \partial c_{\delta} \partial c_{\beta} \partial c^{\dagger}_{\gamma}} \right) c^{\dagger}_{\alpha} c_{\beta} c^{\dagger}_{\gamma} c_{\delta}: \\
&=:c^{\dagger}_{\alpha} c_{\beta} c^{\dagger}_{\gamma} c_{\delta}: + G_{\alpha \delta} :c_{\beta} c^{\dagger}_{\gamma}: + \tilde{G}_{\beta \gamma} :c^{\dagger}_{\alpha} c_{\delta}: + G_{\alpha \delta} \tilde{G}_{\beta \gamma}.
\end{align}
So the commutation relation is given by:
\begin{align}
[:c^{\dagger}_{\alpha} c_{\beta}:, :c^{\dagger}_{\gamma} c_{\delta}:] &= -(G_{\alpha \delta}+\tilde{G}_{\alpha \delta}) :c_{\beta} c^{\dagger}_{\gamma}: + (\tilde{G}_{\beta \gamma}+G_{\gamma \beta}) :c^{\dagger}_{\alpha} c_{\delta}: + (G_{\alpha \delta} \tilde{G}_{\beta \gamma} - G_{\gamma \beta} \tilde{G}_{\delta \alpha}) \\
&=  \delta_{\beta \gamma} :c^{\dagger}_{\alpha} c_{\delta} :- \delta_{\alpha \delta} :c^{\dagger}_{\gamma} c_{\beta} : + (G_{\alpha \delta} \tilde{G}_{\beta \gamma} - G_{\gamma \beta} \tilde{G}_{\delta \alpha}),
\end{align}
which is just the regular commutator plus a constant. The higher-order terms can be calculated similarly, albeit with a much larger number of terms.

\pagebreak
\section{Flow Equation Details}
\subsection{Calculating the flow}
\begin{figure}[h!]
\begin{center}
\includegraphics[width= \linewidth]{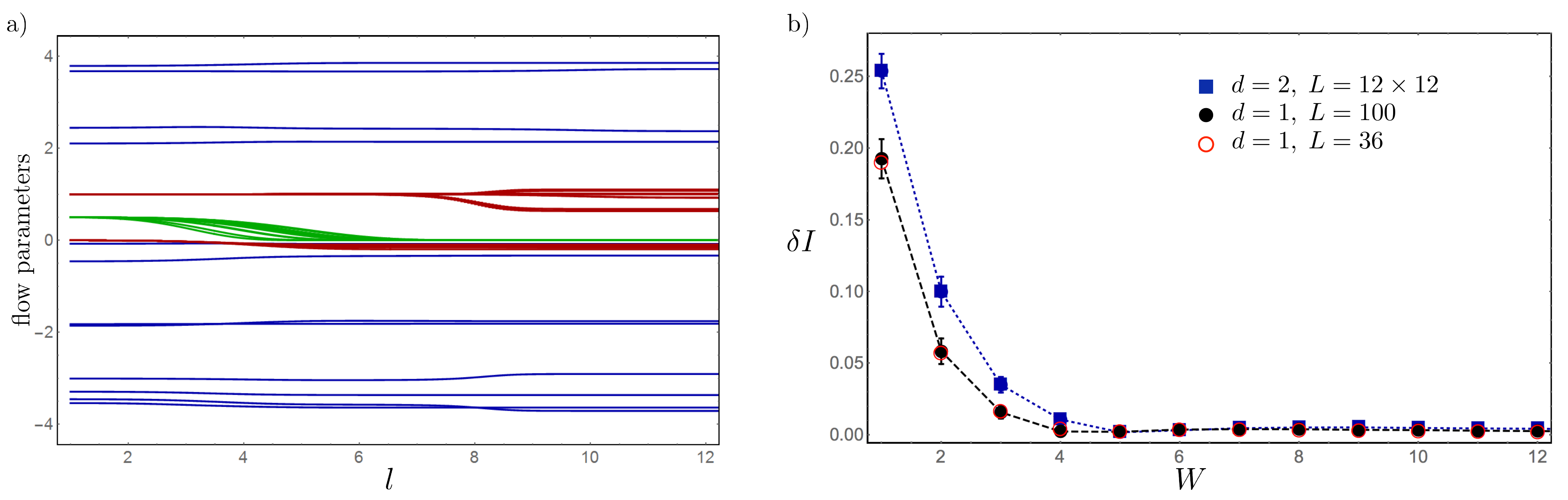}
\caption{a) Typical flow of a representative sample of parameters from a chain of length $L=36$ with $W=4$. The off-diagonal elements $J_{ij}$ are shown in green and all decay essentially exponentially in the later stages of the flow. The diagonal $h_i$ elements are shown in blue, and the interaction terms $\Delta_{ij}$ are shown in red. Note that some of the $\Delta_{ij}$ terms which are initially zero become non-zero at the end of the flow. b) The flow invariant $\delta I$ as defined in the text, showing how the flow is essentially closed for large disorder, but for small disorder the truncated terms carry significant weight and the resulting error can be quite large. This was calculated on chains of length $L=36$ and $L=100$ in $d=1$ and on a grid of size $L=12 \times12$ in $d=2$, with both averaged over $N_s = 100$ disorder realisations.}
\label{fig.one}
\end{center}
\end{figure}

The explicit form of the generator is:
\begin{align}
&\eta(l) = \eta_{free}(l) + \eta_{int}(l) + \eta_{no}(l), \\
&\eta_{free}(l) = \sum_{ij} J_{ij} (h_i -h_j) (:c^{\dagger}_i c_{j}: - :c^{\dagger}_j c_i:), \\
&\eta_{int}(l) = 2 \sum_{ijk} J_{ij} (\Delta_{ik}-\Delta_{jk}) :c^{\dagger}_k c_k(c^{\dagger}_i c_j-c^{\dagger}_j c_i):, \\
& \eta_{no}(l) = \sum_{ijlm} \Delta_{ij} J_{lm} \left[ 2(G_{jm} \tilde{G}_{jl} - G_{lj} \tilde{G}_{mj}) :c^{\dagger}_i c_{i} : + (G_{jm} \tilde{G}_{il} - G_{li} \tilde{G}_{mj}) (:c^{\dagger}_i c_{j}:  - :c^{\dagger}_j c_{i}: ) \right],
\end{align} 
where the first term is identical to the free system, the second term is due to the interactions and the third is the explicit normal-ordering correction to the generator.
To obtain the flow equations shown in the main text, we commute this generator with our ansatz for the running form of the Hamiltonian. We specify to a product state, where all expectation values of the following form vanish:
\begin{align}
G_{\alpha \delta} \tilde{G}_{\beta \gamma} - G_{\gamma \beta} \tilde{G}_{\delta \alpha} = (\delta_{\alpha \delta} \delta_{\beta \gamma} - \delta_{\gamma \beta} \delta_{\delta \alpha}) \langle n_{\alpha} \rangle \langle n_{\beta} \rangle = 0,
\end{align}
which simplifies the resulting equations considerably. In particular, the normal-ordering correction to the generator itself vanishes. The flow equations are then obtained from:
\begin{align}
\frac{\ud \mathcal{H}}{\ud l} &= [\eta(l), \mathcal{H}(l)] = [\eta_{free}(l),\mathcal{H}_0(l)] + [\eta_{free}(l),V(l)] + [\eta_{int}(l),\mathcal{H}_0(l)] + [\eta_{int}(l),V(l)],
\end{align}
which, after using the above product state simplification and discarding all newly-generated terms of quartic order and higher, leads to the final flow equation shown in the main text. A typical flow is shown in Fig. \ref{fig.one}a).

\subsection{Flow invariant}
To quantify the effect of the truncation which we employ, following Ref. \onlinecite{Monthus16} we can make use of the second invariant of the flow, defined as:
\begin{align}
I(l) = \sum_i h_i^2(l) + \sum_{ij}\left(  J^2_{ij}(l) + \Delta^2_{ij}(l) \right) ,
\end{align}
which is a property conserved by the exact unitary flow $U(l)$ and acts as a measure of how much is `lost' by truncating the Hamiltonian. We define the difference $\delta I$ between the initial and final Hamiltonians as follows:
\begin{align}
\delta I = \frac{|I(0)-I(\infty)|}{\frac12 (I(0)+I(\infty))}.
\end{align}
Plotting against disorder, as shown in Fig. \ref{fig.one}b), we see that for large disorder the resulting error remains small and under control, while in the delocalized phase our error becomes quite large. There is no significant dependence of $\delta I$ on system size. The dramatic breakdown of the conservation of the flow invariant for $d=1$ in the vicinity of disorder strength $W \sim 3$ is suggestive of a phase transition, however identifying a transition based on the breakdown of our method is tenuous at best. In $d=2$ the error is larger, consistent with our results in the main text showing that the system is less localized in two dimensions.

\section{Comparison with Exact diagonalization}
\subsection{Spectra}
\begin{figure}[h!]
\begin{center}
\includegraphics[width= \linewidth]{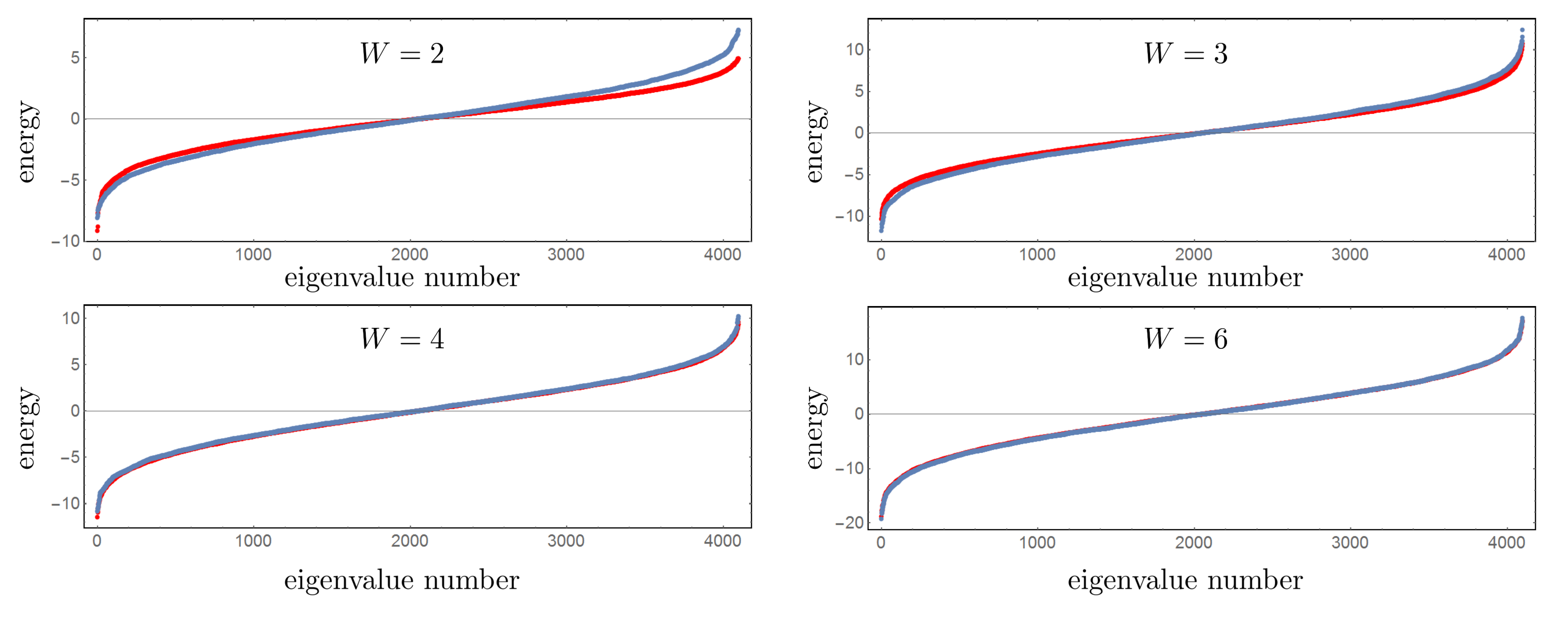}
\caption{Comparison of spectra for a chain of length $L=12$ - ED (red) vs flow equations (blue) at $W=2,3,4,6$. As we increase the disorder and cross the localization transition, the two methods give almost identical results.}
\label{fig.edcomp}
\end{center}
\end{figure}

From the diagonal Hamiltonian, we can extract the full spectrum and compare with exact diagonalization data (shown in Fig. \ref{fig.edcomp}), obtained with the package QuSpin~\cite{Weinberg+17}. We find that in the delocalized regime (small disorder) the spectra visibly differ. The agreement improves as we increase the disorder: deep in the MBL phase the two methods give almost identical results.

\pagebreak
\subsection{Level statistics}
In addition, we can also gain some insight from the level statistics. It was previously established that the level-spacing statistics differ in the localized and delocalized phases. If we consider the ratio of adjacent energy gaps $r_n$ given by:
\begin{align}
\delta_n &= |E_n - E_{n+1}|, \\
r_n &= \textrm{min}(\delta_n,\delta_{n+1})/\textrm{max}(\delta_n,\delta_{n+1}),
\end{align}
it has been shown~\cite{OganesyanHusePRB07} that the average of $r$ tends to $\sim0.53$ in the delocalized phase (corresponding to Wigner-Dyson level statistics), and $\sim0.39$ in the localized phase (corresponding to Poisson level statistics). In exact diagonalization (ED), one obtains this quantity by obtaining the spectrum for a fixed filling fraction (here we use $1/2$), sorting the spectrum into ascending order and calculating the above expressions. By extracting this quantity from both the flow equation framework and from ED, we find that the flow equation result yields Poisson-distributed levels at all disorder strengths, which confirms that this approach is accurate in the MBL phase but misses the transition in the delocalized regime. Although not shown here, the same analysis was performed for the $d=2$ system of size $L = 12 \times 12$ and again we find a value of $\overline{r} \sim 0.39$ for all disorder strengths, consistent with Poisson level statistics. 
\begin{figure}[h!]
\begin{center}
\includegraphics[width= 0.8\linewidth]{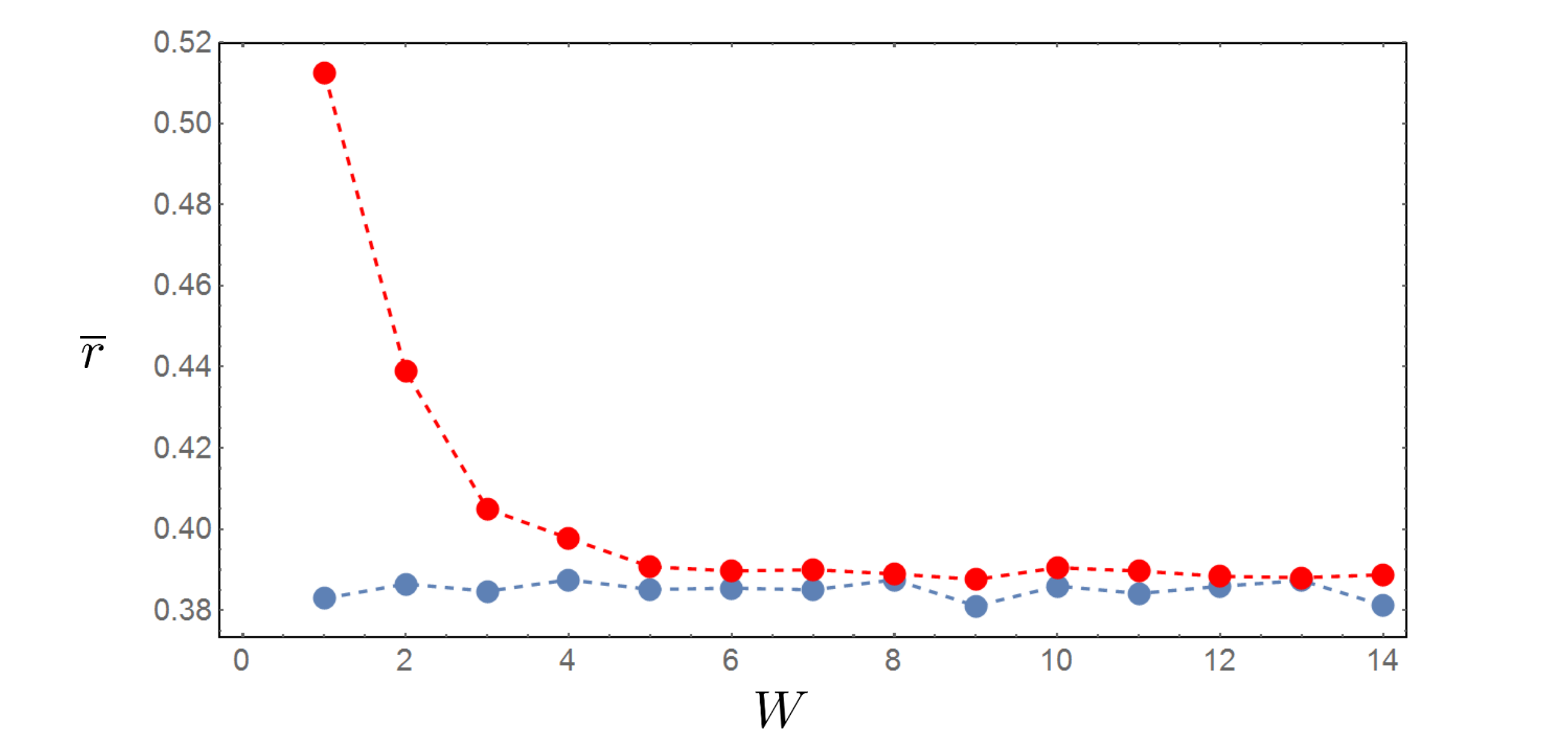}
\caption{Comparison of the level statistics from ED (red) and the flow equation method (blue) for a chain of length $L=12$ at half-filling. The ED correctly reproduces the change in level statistics from Poisson to Wigner-Dyson, whereas the flow equation method converges to Poisson statistics at all disorder strengths.}
\end{center}
\end{figure}

\pagebreak
\section{Full Probability Distribution of the couplings}

In addition to extracting a localization length from the averaged values of the $l$-bit couplings as in the main text, we can also take a look at the full probability distribution functions of the $l$-bit couplings $\Delta_{ij}$ to see how they behave.

To do this, we take a chain of length $L=64$ and sample 1,000 realisations of the random disorder at a variety of disorder strengths. We then take the full distribution of the magnitude of the $\Delta_{ij}$ couplings at different ranges $r = |i-j|$ from across all sampled disorder realisations and plot the probability distribution function (PDF). Note that we have plotted this on a logarithmic scale, rescaled both axes and offset the plots for clarity. We do not consider ranges longer then $r=24$ - the system is localized on length scales much shorter than this and the couplings at this length scale are already so small as to be below reasonable numerical accuracy.

At all disorder strengths $d$ and ranges $r$ we find approxmate power-law decay, with slight deviations. We see no signs of the scale-invariant or white-noise-like distributions seen for weak disorder found in Ref. \onlinecite{Pekker+16}, though this is not surprising as our method is not valid for weak disorder and always converges to a localized phase. At the large disorder strengths, we do indeed see power-law behaviour consistent with Ref. \onlinecite{Pekker+16}. At small disorder strengths, due to the limitations of our method the PDF displays the same localized behaviour. Our results differ from the discrete displacement transform studied in Ref. \onlinecite{Rademaker+17}. Although we also see that the tails of the distribution deviate from power-law at the longest lengthscales and largest disorder we consider, we do not find the same transition to exponential decay. In the strong disorder limit where our method is most accurate, and at small length scales where we can be sure we are within numerical accuracy, both Refs. \onlinecite{Pekker+16,Rademaker+17} and ourselves find power-law decay of the couplings.

\begin{figure}[h!]
\begin{center}
\includegraphics[width= 0.75\linewidth]{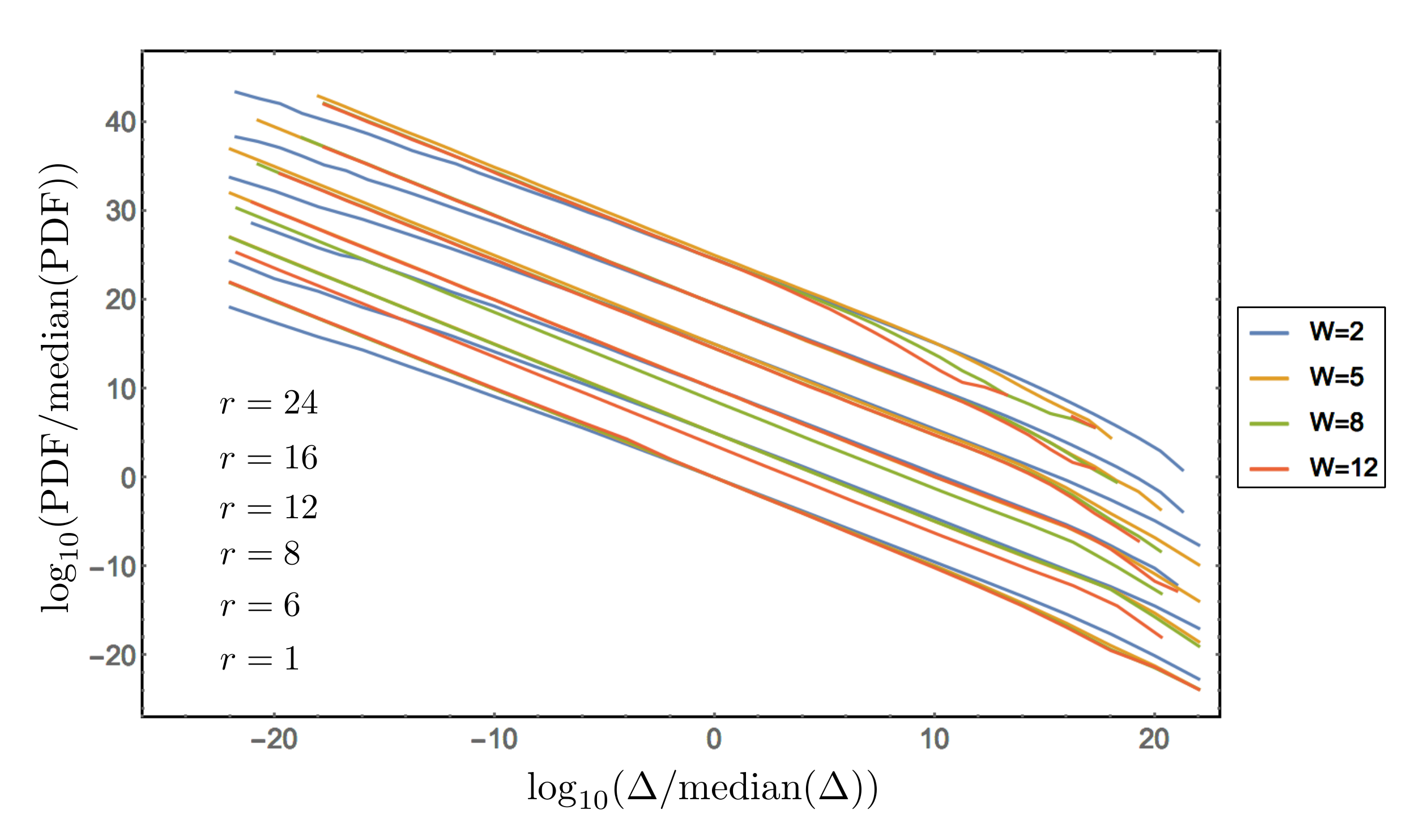}
\caption{Probability distribution functions (PDFs) of the rescaled $l$-bit couplings for a variety of disorder strengths $W$ and ranges $r=|i-j|$. Both axes have been have been normalised and the data vertically offset for clarity.}
\label{fig.PDFs}
\end{center}
\end{figure}

